\documentclass[11pt]{article}
\usepackage[utf8]{inputenc}
\usepackage[T1]{fontenc}
\usepackage{newtxtext,newtxmath}        
\usepackage[margin=1in]{geometry}       
\usepackage{setspace}
\usepackage{amsmath}
\usepackage{graphicx}
\usepackage[superscript,biblabel]{cite} 
\usepackage{authblk}
\usepackage{url}
\usepackage{tabularx}
\usepackage{booktabs}
\usepackage[labelfont=bf, font=small]{caption}

\renewenvironment{abstract}{\par\noindent\bfseries}{\par\medskip}

\title{From picometre displacements to a three-dimensional lattice gas of polar skyrmions}

\begin{document}
\author[1,2,*]{Joohee Bang}
\author[3]{Nives Strkalj}
\author[4]{Hugo Aramberri}
\author[2]{Stephan Rosenkranz}
\author[1,5]{Arkadiy Simonov}
\author[1]{Morgan Trassin}
\author[1,*]{Thomas Weber}

\affil[1]{Department of Materials, ETH Z\"urich, Z\"urich, Switzerland}
\affil[2]{Materials Science Division, Argonne National Laboratory, Lemont, IL, USA}
\affil[3]{Center for Advanced Laser techniques, Institute of physics, Zagreb, Croatia}
\affil[4]{Smart Materials Unit, Luxembourg Institute of Science and Technology, Esch/Alzette, Luxembourg}
\affil[5]{Department of Chemistry, Aarhus University, Aarhus, Denmark}
\affil[*]{Corresponding authors. Emails: joohee.bang@anl.gov, thomas.weber@mat.ethz.ch}

\date{}

\maketitle

\begin{abstract}
Polar skyrmions, the electric analogues of magnetic skyrmions, form arrays in PbTiO$_3$/SrTiO$_3$ superlattices that are known to be only short-range ordered, but how this disordered ensemble is organized in three dimensions, and how its microscopic structure underlies this organization, have remained largely inaccessible. Here, using three-dimensional difference pair distribution function (3D-$\Delta$PDF) analysis of X-ray diffuse scattering, we show that the polar skyrmion bubbles form a three-dimensional lattice gas: they occupy discrete sites, but their order extends over only a few neighbours and is two to three times stronger across the SrTiO$_3$ spacers than within a layer. This interlayer coupling is consistent with the polarization of the bubbles extending into the spacers. Within each bubble, the polarization rotates in a predominantly N\'eel-type manner, and Pb and Ti rotate together with comparable picometre displacements, demonstrating that both $A$- and $B$-site cations take part in the polar distortion.

\end{abstract}

\setcounter{secnumdepth}{0}

Polar topologies, including vortices~\cite{Naumov2004,AguadoPuente2008,Tang2015,yadav2016observation}, merons~\cite{wang2020polar,Shao2023} and skyrmions~\cite{Nahas2015,Goncalves2019,das2019observation}, have been predicted and stabilized in ferroelectrics as electric analogues of their magnetic counterparts~\cite{Muhlbauer2009,Yu2010,Nagaosa2013}. They emerge from a balance of electrostatic, elastic and gradient energies~\cite{Junquera2023}, are only a few nanometres across and are electrically switchable, making them candidates for low-energy, high-density, non-volatile memory~\cite{han2022high}.

PbTiO$_3$/SrTiO$_3$ superlattices are the prototypical platform for polar topologies: polar vortices and skyrmions form in the ferroelectric PbTiO$_3$ layers, separated by paraelectric SrTiO$_3$ spacers~\cite{yadav2016observation,das2019observation,Hong2017,Damodaran2017,Shafer2018,Hong2018,Das2021,Gong2023,Junquera2023}. Real-space imaging has shown their polarization rotation and their arrangement over local fields of view~\cite{yadav2016observation,das2019observation,Das2021,Gong2023,Shao2023}, and imaging and diffraction have shown that polar skyrmion arrays are only short-range ordered~\cite{das2019observation,Shao2023}.

What has remained largely inaccessible~\cite{Shao2023} is how this disordered ensemble is organized: over which distances and along which directions order persists. For magnetic skyrmions, the collective state is known to range from crystalline through hexatic to liquid, reflecting how skyrmions interact with one another and with the host lattice~\cite{Huang2020,Karube2016,Nagaosa2013,Fert2017}. For polar skyrmions, the collective state has so far been examined within individual layers and over local fields of view. In PbTiO$_3$/SrTiO$_3$ superlattices, however, it must be resolved in three dimensions, because coupling through the spacers decides whether polar skyrmions in neighbouring layers stack into columns or behave as independent sheets. This collective organization is in turn tied to the internal structure of each bubble: its shape, the character of its domain walls and the displacements of the individual cations, including whether the $A$- and $B$-site cations rotate together. Understanding polar skyrmions therefore requires resolving both levels together, in three dimensions and within the buried film.

Here, we address these questions using three-dimensional difference pair distribution function (3D-$\Delta$PDF) analysis~\cite{weber2012three, Bang2025} of X-ray diffuse scattering from PbTiO$_3$/SrTiO$_3$ superlattices. The 3D-$\Delta$PDF maps how pairs of atoms deviate from the average structure along every direction, including through the film thickness, and references the displacements of each cation species to the undistorted lattice of the film itself. In non-destructive measurements of as-grown films, we show that the polar skyrmion bubbles form a three-dimensional lattice gas, in the sense of the lattice-gas model of statistical physics~\cite{LeeYang1952}: the bubble centres occupy discrete sites of the perovskite lattice, and the order between bubbles extends over only a few neighbours, without long-range order. This order is stronger across the SrTiO$_3$ spacers than within a layer, and the same hierarchy is found in superlattices with different layer thicknesses. Within each bubble, the polarization rotates in a predominantly N\'eel-type manner across the domain walls, and Pb and Ti rotate together with comparable picometre amplitudes.

\section{Diffuse scattering analyzed with 3D-$\Delta$PDF}
Bragg reflections and crystal truncation rods (CTRs) encode the average structure of the crystal, whereas diffuse scattering arises from local deviations from it, here the polar skyrmion bubbles. To capture this weak diffuse signal throughout a large volume of reciprocal space, we measured the three-dimensional X-ray scattering of PbTiO$_3$/SrTiO$_3$ superlattices using high-energy ultra-small-angle grazing-incidence diffraction~\cite{Bang2025}.

We focus on the (PbTiO$_3$)$_n$/(SrTiO$_3$)$_m$ superlattice with $n=3$ and $m=5$, where $n$ and $m$ are the layer thicknesses in unit cells. With three-unit-cell PbTiO$_3$ layers, this superlattice lies near the predicted limit of polar-skyrmion stability, beyond which the depolarized PbTiO$_3$ becomes paraelectric~\cite{WangChen2024}, although polar skyrmion bubbles have been observed in layers as thin as two unit cells~\cite{Gong2023}. Superlattices with $(n, m) = (5, 2)$ and $(7, 7)$ were measured for comparison.

\begin{figure}[h]
\centering
\includegraphics[width=\textwidth]{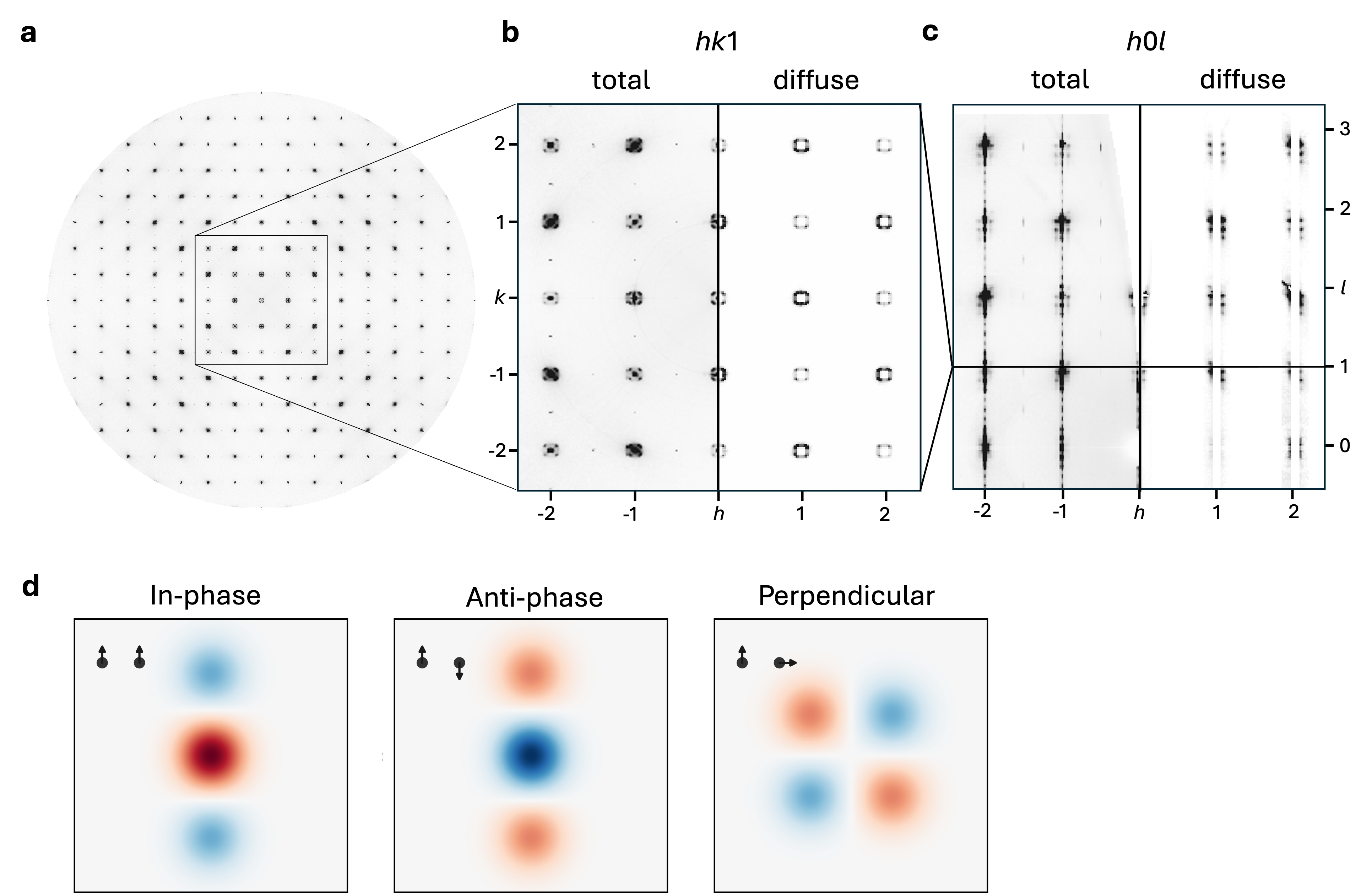}
\caption{\textbf{Three-dimensional X-ray diffuse scattering from polar skyrmion bubbles.}
\textbf{a}, $hk1$ layer of the $n=3$, $m=5$ PbTiO$_3$/SrTiO$_3$ superlattice at room temperature, reconstructed from the symmetry-expanded three-dimensional dataset. Each crystal truncation rod (CTR) is surrounded by a square-shaped diffuse ring, whose radius is set by the size of the polar skyrmion bubbles.
\textbf{b}, Magnified region of \textbf{a} (box).
\textbf{c}, $h0l$ layer, in which the superlattice satellites around the CTRs are cross-sections of the diffuse rings. In \textbf{b} and \textbf{c}, total scattering (left half) is shown next to the isolated diffuse scattering (right half), from which the CTRs and the weak streaks at half-integer $h,k$ arising from octahedral tilts in the SrRuO$_3$ buffer layer have been removed. Vertical gaps in \textbf{c} mark the removed CTRs; its top-left region is inaccessible in the grazing-incidence geometry. \textbf{d}, Schematic 3D-$\Delta$PDF signatures of a single atom pair (arrows: displacements of the two atoms). In-phase displacements give a positive Mexican-hat profile, anti-phase displacements a negative one, and perpendicular displacements a four-lobed checkerboard pattern. Red and blue denote positive and negative $\Delta$PDF density.}
\label{fig:fig1}
\end{figure}

In the reconstructed data, every CTR is surrounded by a set of square-shaped diffuse rings (Fig.~\ref{fig:fig1}a,b), whose radius is set by the size of the bubbles and whose fourfold intensity modulation around the ring reflects their in-plane shape and arrangement; the superlattice satellites around the CTRs in the $h0l$ plane are cross-sections of these rings (Fig.~\ref{fig:fig1}c). Fourier transformation of the diffuse scattering, after removal of the CTRs and of scattering unrelated to the bubbles (Methods), yields the 3D-$\Delta$PDF~\cite{weber2012three}: a map of how pairs of atoms separated by the interatomic vector $\mathbf{u}=(u,v,w)$ deviate from the average structure, here the centrosymmetric perovskite. Positive (red) values indicate that such pairs occur more often than in the average structure, and negative (blue) values less often, weighted by the product of the scattering factors of the two atoms.

The maps can be read directly in terms of coupled atomic displacements (Fig.~\ref{fig:fig1}d). Each atom pair contributes a feature at its interatomic vector, elongated along the joint displacement of the two atoms. In-phase displacements give a positive maximum at the interatomic vector, flanked by two negative lobes along the displacement direction (a positive Mexican-hat profile); antiphase displacements give the inverted profile, a negative minimum flanked by two positive lobes (a negative Mexican-hat profile). Pairs within one polar domain therefore give positive Mexican hats, and pairs spanning oppositely polarized domains give negative ones. If one atom is displaced out of plane and the other in plane, as for an atom inside a domain paired with one in a domain wall, the feature instead consists of four lobes of alternating sign arranged in a checkerboard pattern. Such four-lobed features signal polarization rotation between the two atoms, linking the 3D-$\Delta$PDF directly to the noncollinear polarization structure.

\section{Collective N\'eel-type rotation of Pb and Ti}
\begin{figure}[h]
\centering
\includegraphics[width=\textwidth]{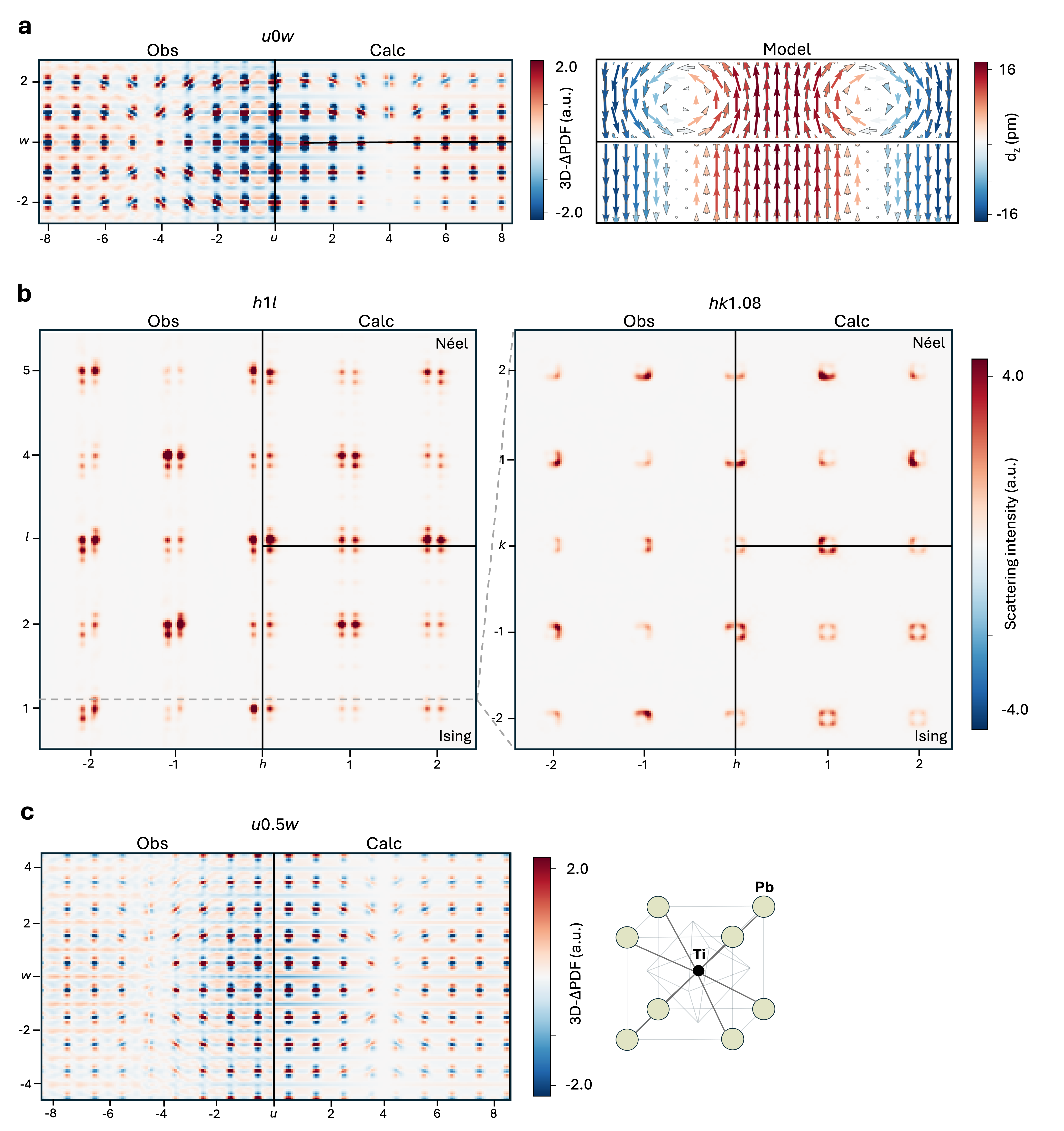}
\caption{\textbf{Collective N\'eel-type polarization rotation of Pb and Ti.} Observed and calculated 3D-$\Delta$PDF sections and diffuse scattering for the $n=3$, $m=5$ superlattice. Red and blue indicate positive and negative 3D-$\Delta$PDF values; $u$, $v$, $w$ are in units of the SrTiO$_3$ lattice parameter. \textbf{a}, $u0w$ section of the observed (left) and calculated (right) 3D-$\Delta$PDF, with calculations from the N\'eel-type (top) and Ising-type (bottom) polarization rotation models. The models are shown as vertical cross-sections on the right; arrows indicate cation displacements, coloured by the out-of-plane component $d_z$. \textbf{b}, Observed (left) and calculated (right) diffuse scattering in the $h1l$ and $hk1.08$ layers for the N\'eel (top) and Ising (bottom) models; the dashed line marks $l = 1.08$. \textbf{c}, $u\tfrac{1}{2}w$ section of the observed (left) and calculated (right, N\'eel model) 3D-$\Delta$PDF, containing the Pb--Ti vectors $\langle i+\tfrac{1}{2}, \tfrac{1}{2}, k+\tfrac{1}{2}\rangle$ shown on the right.}
\label{fig:fig2}
\end{figure}

We show how the 3D-$\Delta$PDF resolves noncollinear arrangements of electric dipoles. Figure~\ref{fig:fig2}a shows the $u0w$ section of the 3D-$\Delta$PDF together with sections calculated for Ising- and N\'eel-type domain-wall models. The strongest features lie at the integer vectors $\langle i, 0, k\rangle$ ($i, k \in \mathbb{Z}$), corresponding to Pb--Pb and Ti--Ti pairs. Each feature is oriented along the correlated displacements of the pair. At short lateral vector lengths, the features are positive Mexican-hat profiles elongated along $w$, arising from pairs within the same domain with parallel out-of-plane displacements. At large lateral vector lengths, they are negative, arising from pairs spanning oppositely polarized domains. The sign reversal is not abrupt. With increasing lateral vector length, the profiles tilt away from $w$, pass through four-lobed checkerboard features, and rotate until the opposite-signed profile along $w$ is recovered. The checkerboard features link a domain to its wall, where the displacements of the two atoms are orthogonal, so the rotation of the features traces the rotation of the polarization across the wall. The Ising model, with only out-of-plane displacements, reproduces the sign reversal but not the rotation, which appears only when the polarization acquires an in-plane component at the wall.

The rotation is predominantly N\'eel-type: in a N\'eel-type wall, the in-plane displacements point radially, parallel to the vectors of pairs along a bubble radius, so the checkerboard features appear in sections containing $w$ and the radial direction, such as $u0w$. In a Bloch-type wall, the in-plane displacements would instead be tangential, perpendicular to these pair vectors, and the checkerboard features would appear in sections perpendicular to $u0w$ rather than within it.

The N\'eel model reproduces the data not only in PDF space but also in reciprocal space (Fig.~\ref{fig:fig2}b). The diffuse rings surrounding the CTRs show a pronounced intensity anisotropy along the scattering vector, with stronger intensity on the low-$|\mathbf{Q}|$ side of each ring. It is visible as unequal intensities in the ring cross-sections in the $h1l$ layer and around the full rings in the $hk1.08$ layer. This anisotropy is reproduced by the N\'eel model, whereas the Ising model yields uniform rings.

The signature of polarization rotation is not confined to the integer vectors but also appears at the half-integer vectors. The $u\tfrac{1}{2}w$ section (Fig.~\ref{fig:fig2}c) contains the vectors $\langle i+\tfrac{1}{2}, \tfrac{1}{2}, k+\tfrac{1}{2}\rangle$, which connect $A$- and $B$-site cations (Fig.~\ref{fig:fig2}c, right) and, within the PbTiO$_3$ layers, are dominated by Pb--Ti pairs. The features evolve as in the $u0w$ section, rotating from positive profiles along $w$ at short in-plane separations to negative profiles at larger separations. Pb and Ti are therefore displaced in phase and rotate together across the domain wall: the Pb sublattice takes part collectively in the polarization rotation and, together with Ti, forms the polar skyrmion at the microscopic level. The refined displacements confirm this: Pb moves by 13~pm out of plane and 6~pm in plane, comparable to the 11~pm and 5~pm of Ti (Supplementary Table~\ref{tab:rms}).

\section{Polar skyrmion lattice gas coupled across the SrTiO$_3$ spacers}

\begin{figure}[h]
\centering
\includegraphics[width=\textwidth]{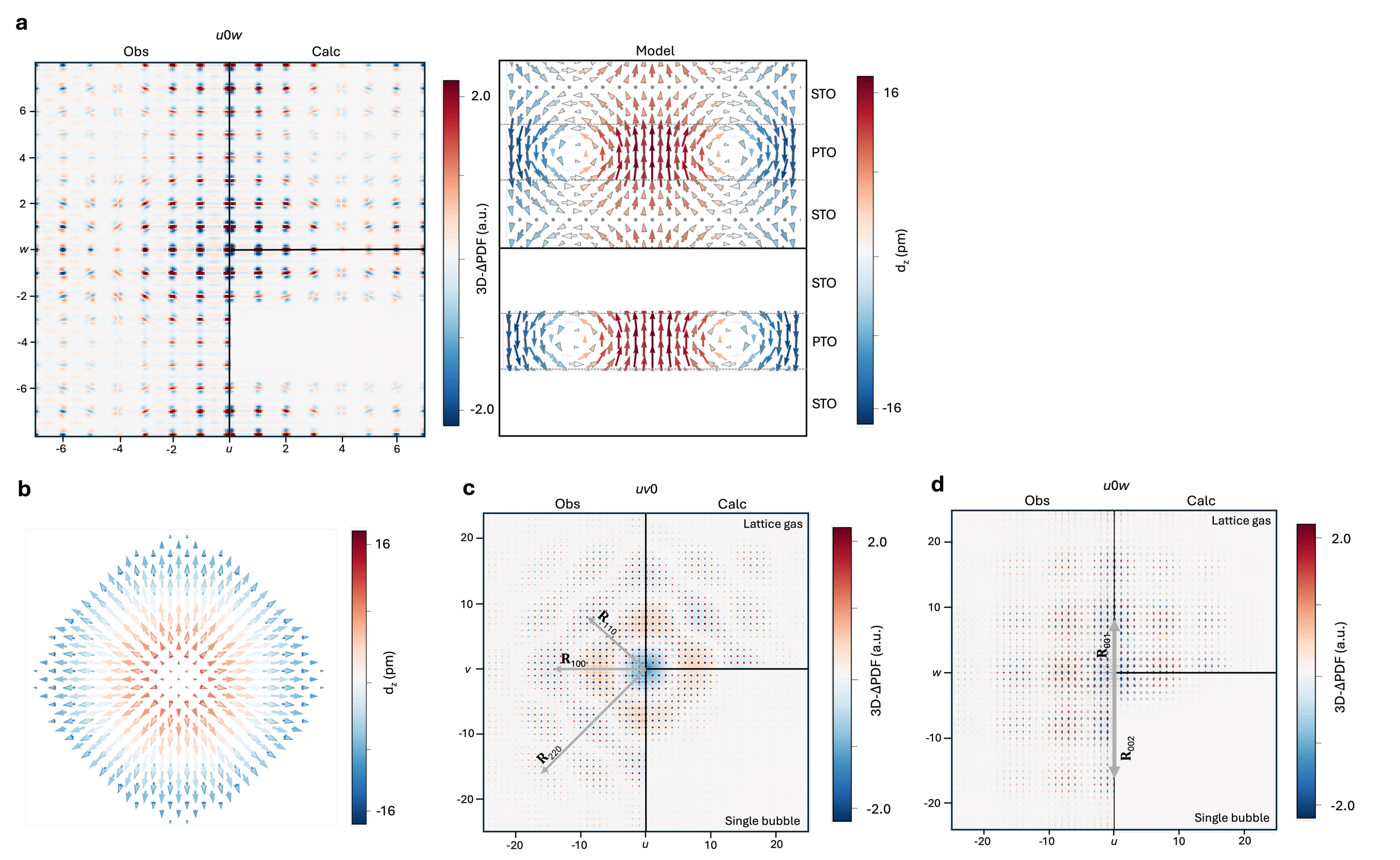}
\caption{\textbf{Polar skyrmion bubbles form a lattice gas coupled across the SrTiO$_3$ spacers.} Observed and calculated 3D-$\Delta$PDF of the $n=3$, $m=5$ superlattice. \textbf{a}, $u0w$ section of the observed (left) and calculated (right) 3D-$\Delta$PDF. The calculated half compares models with displacements extending into the SrTiO$_3$ spacers (top) and confined to the PbTiO$_3$ layers (bottom), shown as vertical cross-sections on the right, with arrows indicating the cation displacements, coloured by their out-of-plane component $d_z$. Features at $3 \le |w| \le 5$, which arise only from pairs involving SrTiO$_3$, are reproduced only when the displacements extend into the spacers. \textbf{b}, Top view of a single refined bubble, with arrows coloured by $d_z$. \textbf{c}, \textbf{d}, $uv0$ (\textbf{c}) and $u0w$ (\textbf{d}) sections of the observed (left) and calculated (right) 3D-$\Delta$PDF over a larger range. The calculated half compares the lattice-gas model (top) with a single bubble (bottom); features beyond the autocorrelation volume of a single bubble arise only from correlations between bubbles. Arrows mark the neighbour vectors $\mathbf{R}$ between bubbles; the refined correlation coefficients are listed in Supplementary Table~\ref{tab:correlations}. Colours and axes as in Fig.~\ref{fig:fig2}.}
\label{fig:fig3}
\end{figure}

In the $n=3$, $m=5$ superlattice, the polar order of the thin PbTiO$_3$ layers is stabilized by screening from the SrTiO$_3$, which itself becomes polarized~\cite{WangChen2024}. Since $w$ counts unit cells along the stacking direction, cation pairs within one PbTiO$_3$ layer have $|w| \le 2$, and with a superlattice period of eight unit cells, pairs in adjacent PbTiO$_3$ layers have $6 \le |w| \le 10$. The window $3 \le |w| \le 5$ therefore contains only pairs involving SrTiO$_3$. The experimental $u0w$ section shows clear features in this window, which are reproduced only when the polar displacements extend into the spacers (Fig.~\ref{fig:fig3}a). In the refined model, Sr and Ti in the spacers are displaced by 7~pm and 5~pm out of plane (Supplementary Table~\ref{tab:rms}). The out-of-plane lattice parameter of the SrTiO$_3$ layers, $c_\mathrm{STO}=3.947$~\AA{}, is expanded by about 1\% relative to bulk SrTiO$_3$ ($c=3.905$~\AA{}), consistent with a polar distortion of the spacers induced by the adjacent PbTiO$_3$ layers.

The same data capture how the bubbles are organized on the mesoscale. Figure~\ref{fig:fig3}b shows a single refined bubble, which has a fourfold-symmetric in-plane cross-section with its corners along $\langle100\rangle$, consistent with the fourfold in-plane symmetry imposed by the cubic substrate. A single bubble contributes to the 3D-$\Delta$PDF only within its own autocorrelation volume. Features beyond this range are absent in the single-bubble calculation and arise exclusively from correlations between bubbles (Fig.~\ref{fig:fig3}c,d). These correlations show that the bubbles form a lattice gas: correlations between bubbles appear at discrete neighbour vectors along $\langle100\rangle$ and $\langle110\rangle$, rather than as an isotropic distribution of separations as in a liquid, but decay rapidly with distance.

We quantify these correlations by refined coefficients $\rho$, ranging from $+1$ for perfect positive to $-1$ for perfect negative correlation, with $0$ for no correlation between neighbouring bubbles (Methods; Supplementary Table~\ref{tab:correlations}). The correlations are strongly anisotropic. Bubbles in adjacent PbTiO$_3$ layers are strongly correlated ($\rho_{001}\approx0.60$), and the correlation decreases to $\rho_{002}\approx0.25$ for the second neighbouring layer, so that the bubbles stack into columnar but still short-range-ordered arrangements across the SrTiO$_3$ spacers (Fig.~\ref{fig:fig3}d). Within a layer, the nearest neighbours along $\langle100\rangle$ and $\langle110\rangle$, at about 55 and 44~\AA{}, are less than half as strongly correlated ($\rho_{100}\approx0.23$, $\rho_{110}\approx0.21$), and the correlation falls to $\rho_{220}\approx0.08$ for the second neighbour along $\langle110\rangle$ (Fig.~\ref{fig:fig3}c). The bubbles are thus correlated two to three times more strongly across the SrTiO$_3$ spacers than within a layer. Across a five-unit-cell spacer, this strong interlayer coupling is consistent with the polarization extending into the SrTiO$_3$ (Fig.~\ref{fig:fig3}a), linking the internal structure of the bubbles to their collective organization. The same anisotropy is found in the other superlattices (Supplementary Fig.~\ref{fig:SI_samples}) and is most pronounced for $n=5$, $m=2$, where bubbles in adjacent layers are as strongly correlated as here ($\rho_{001}\approx0.62$) but show almost no in-plane correlation ($\rho_{100}\approx0.03$).

\section{Methods}
\subsection{Sample preparation}

(PbTiO$_3$)$_n$/(SrTiO$_3$)$_m$ superlattices with $(n, m) = (3, 5)$, $(5, 2)$ and $(7, 7)$ were grown by pulsed-laser deposition on single-crystalline (001)-oriented SrTiO$_3$ substrates ($5 \times 5$~mm$^2$; Crystec GmbH), following the procedure described in ref.~\cite{Bang2025}. PbTiO$_3$ ($a = 3.904$~\AA) and SrTiO$_3$ ($a = 3.905$~\AA) are closely lattice matched in plane. A ten-unit-cell-thick SrRuO$_3$ buffer layer was deposited before the superlattice. Each superlattice consisted of 12 bilayers, starting with a PbTiO$_3$ layer and terminated by an additional PbTiO$_3$ capping layer. Films were deposited using a KrF excimer laser (wavelength 248~nm, pulse duration 20~ns), and two-dimensional layer-by-layer growth was monitored in situ by reflection high-energy electron diffraction with unit-cell precision. The substrate temperature was 700~$^\circ$C for SrRuO$_3$ and 550~$^\circ$C for PbTiO$_3$ and SrTiO$_3$, at an oxygen partial pressure of 0.12~mbar for all layers. The laser fluence was 0.7~J~cm$^{-2}$ for SrRuO$_3$ and 0.8~J~cm$^{-2}$ for PbTiO$_3$ and SrTiO$_3$, with repetition rates of 2~Hz (SrRuO$_3$), 4~Hz (PbTiO$_3$) and 2~Hz (SrTiO$_3$).

\subsection{Three-dimensional X-ray diffuse scattering }
Three-dimensional diffuse scattering data of the thin films were collected by high-energy grazing-incidence X-ray diffraction on beamline P07 at PETRA III, DESY (Hamburg, Germany), as described in detail in ref.~\cite{Bang2025}. The X-ray energy was 74~keV ($\lambda = 0.168$~\AA). The beam was focused onto the sample to a size of 30~$\mu$m (horizontal) $\times$ 2~$\mu$m (vertical) and a convergence angle of about 0.003$^\circ$. The sample surface was aligned parallel to the beam to within 0.004$^\circ$, and data were collected at an incidence angle of 0.03$^\circ$ while the sample was rotated by 200$^\circ$ about its surface normal. Diffraction patterns were recorded at room temperature with a PILATUS3 X CdTe 2M area detector at a sample-to-detector distance of 493.7~mm. Each scan was repeated with the detector shifted by the width of the gaps between its modules, to fill in the missing regions. Orientation matrices were determined with XDS~\cite{Kabsch2010} from additional measurements with a strongly attenuated beam, in which the substrate reflections were unsaturated and the film signal was suppressed. Reciprocal-space volumes were reconstructed with Meerkat~\cite{Simonov2015}. All reciprocal-space coordinates refer to the cubic SrTiO$_3$ substrate ($a = 3.905$~\AA).

\subsection{Data processing and 3D-$\Delta$PDF calculation}

The reconstructed reciprocal-space volumes were processed as described in ref.~\cite{Bang2025}. Detector halos around intense substrate reflections were masked. Because the intensity across frames was modulated by small variations of the grazing-incidence geometry, the asymmetric unit with the strongest signal was symmetry-expanded assuming 4/$mmm$ Laue symmetry, consistent with the $p4mm$ layer symmetry of the substrate surface. To isolate the diffuse scattering from the superlattice, the crystal truncation rods, which carry the average structure of the film and substrate, were removed, together with the blind region near $h = k = 0$; regions between the diffuse rings were set to zero. The 3D-$\Delta$PDF was obtained by Fourier transformation of the processed diffuse scattering~\cite{weber2012three}. The same masks were applied to the calculated diffuse scattering before Fourier transformation, so that observed and calculated 3D-$\Delta$PDFs contain identical truncation effects.

\subsection{Structural model of polar skyrmion bubbles}

The model describes spatially varying displacements of the $A$-site (Pb, Sr) and $B$-site (Ti) cations in PbTiO$_3$ and SrTiO$_3$, with coupled in-plane and out-of-plane components; oxygen is not included. The cations are displaced predominantly out of plane, with opposite polarization at the centre and at the perimeter of the bubble. At the domain wall, in-plane components, proportional to the gradient of the out-of-plane displacement and to the height above the centre of the PbTiO$_3$ layer, rotate the polarization in a N\'eel-type manner. Setting these in-plane components to zero yields the Ising-type model used for comparison in Fig.~\ref{fig:fig2}. A single bubble is defined by its size and shape, the width of its domain wall, the displacement amplitudes of each element and the extent of the displacement field into the adjacent SrTiO$_3$ layers. 

Bubbles $S$ and their counterparts $S'$, obtained by mirroring $S$ through the plane parallel to the substrate that contains the bubble centre, are assumed to occur with equal probability, so that the average structure remains centrosymmetric. The diffuse scattering of a single bubble is then
\begin{equation}
I^{\mathrm{single}}_{\mathrm{diff}}(\mathbf{h}) = \left| \mathcal{F}\{S(\mathbf{x}) - S'(\mathbf{x})\} \right|^2,
\end{equation}
where $\mathbf{x} = (x, y, z)$ are real-space coordinates and $\mathbf{h}$ is the scattering vector, and the corresponding 3D-$\Delta$PDF is
\begin{equation}
P^{\mathrm{single}}_{\Delta}(\mathbf{u}) = \mathcal{F}^{-1}\{I^{\mathrm{single}}_{\mathrm{diff}}(\mathbf{h})\}.
\end{equation}
In a real sample, bubbles vary in size, shape and internal structure, so that the displacements of distant atoms within the same bubble are less correlated than in a model of identical bubbles. We account for this by multiplying $P^{\mathrm{single}}_{\Delta}$ by a function $D(\mathbf{u})$ that damps the calculated signal with distance along the in-plane and out-of-plane directions. A flat $D(\mathbf{u})$ corresponds to a homogeneous set of bubbles, whereas strong damping indicates a wide variety of bubble sizes, shapes and internal structures. Thermal and static displacements of individual atoms are described by anisotropic atomic displacement parameters.

The organization of the bubbles is introduced through coefficients $\rho_{\mathbf{R}} \in [-1, +1]$ for neighbouring bubbles separated by $\mathbf{R}$, within a layer and across the SrTiO$_3$ spacers:
\begin{equation}
P^{\mathrm{all}}_{\Delta}(\mathbf{u}) = P^{\mathrm{single}}_{\Delta}(\mathbf{u}) + \sum_{\mathbf{R}} \rho_{\mathbf{R}}\, P^{\mathrm{single}}_{\Delta}(\mathbf{u} - \mathbf{R}).
\end{equation}
Positive $\rho_{\mathbf{R}}$ indicates a preference for neighbouring bubbles of the same polarity, negative $\rho_{\mathbf{R}}$ a preference for opposite polarity, and $\rho_{\mathbf{R}} = 0$ no preference, in which case such pairs give no signal. In-plane neighbours lie on lattice vectors of the perovskite lattice; out-of-plane neighbours are separated by multiples of the superlattice period.

\subsection{Refinement}

The model was refined against the experimental 3D-$\Delta$PDF over the full measured volume of $800^3$ voxels, with the same masks applied to the observed and calculated data. The parameters fall into two groups. Displacement amplitudes, lattice parameters, damping parameters and correlation coefficients change the calculated 3D-$\Delta$PDF smoothly and were refined by least squares. Parameters that determine which unit cells belong to a bubble (its radii and shape exponents), as well as integer parameters such as the neighbour distances and the extent of the bubble into the SrTiO$_3$, change the model in discrete steps, so least-squares gradients are not defined for them. These were determined by evaluating candidate values and by differential evolution~\cite{Storn1997}, a global optimization method previously applied to disorder models refined against diffuse scattering~\cite{Weber2002}. Refinements started from random and from physically motivated initial values; most converged to the same solution, and the remaining ones either differed only in strongly correlated parameters or gave clearly higher $R$ values.

\section{Discussion}

We have shown that the three-dimensional organization of polar topologies, from the correlations between neighbouring bubbles within and across layers down to the cation displacements forming each bubble, can be resolved with picometre sensitivity in a single non-destructive measurement of the as-grown film. In PbTiO$_3$/SrTiO$_3$ superlattices, the polar skyrmion bubbles form a three-dimensional lattice gas, correlated two to three times more strongly across the SrTiO$_3$ spacers than within a layer, consistent with their polarization extending into the spacers. Within each bubble, the domain walls are predominantly N\'eel-type, and Pb and Ti rotate collectively across them. A single structural model reproduces the measured 3D-$\Delta$PDF from pairs of neighbouring bubbles to interatomic pairs.

The approach applies broadly to ferroic topologies in epitaxial films and heterostructures. It distinguishes domain-wall types, accesses structures only a few nanometres across buried within a heterostructure, and is compatible with non-ambient conditions; the in-plane ordering of the bubbles can, for example, be followed in situ up to the paraelectric phase and back (Supplementary Fig.~\ref{fig:SI_temperature}). This opens the way to structural studies of vortices, merons and skyrmions under electric field, illumination and temperature. Because the response of such systems depends not only on the structure of individual topologies but also on how they are organized and coupled in three dimensions, resolving this organization quantitatively provides a structural basis for understanding and controlling emergent functionality in ferroic materials.

\bibliographystyle{unsrt}   
\bibliography{sn-bibliography}       

\clearpage   

\section*{Acknowledgements}
We thank Martin F. Sarott and Ipek Efe for the assistance with the pulsed-laser deposition growth of the superlattices. This work was supported by the Swiss National Science Foundation (grant no.~20021\_188663). Work at Argonne National Laboratory was supported by the U.S. Department of Energy, Office of Science, Basic Energy Sciences, Materials Sciences and Engineering Division. We acknowledge DESY (Hamburg, Germany), a member of the Helmholtz Association HGF, for the provision of experimental facilities. All X-ray diffuse scattering experiments were carried out at beamline P07 of PETRA III, and we would like to thank Ann-Christin Dippel, Olof Gutowski, and Martin von Zimmermann for assistance in using beamline P07. We gratefully acknowledge the computing resources provided by the Laboratory Computing Resource Center at Argonne National Laboratory.

\section*{Author contributions}
J.B. and T.W. conceived the study. J.B. grew the samples under the supervision of M.T., with N.S. contributing to the optimization of the growth conditions. J.B. and T.W. designed and performed the X-ray diffuse scattering experiments and analyzed the data. J.B., T.W. and A.S. developed the structural modelling and refinement code, and H.A. contributed to the modelling. S.R. provided the computational resources for the large-scale structural refinements. M.T., N.S. and H.A. contributed to the interpretation of the results in the context of ferroelectric polar textures. J.B. and T.W. wrote the manuscript with contributions from all authors.

\section*{Competing interests}
The authors declare no competing interests.

\section*{Data availability}
The full three-dimensional reciprocal-space reconstructions, the observed and calculated 3D-$\Delta$PDF volumes and the refined structural models are available from the corresponding authors upon request and will be deposited in a public repository upon publication.

\section*{Code availability}
The code used for the structural modelling and refinements is available from the corresponding authors upon request and will be deposited in a public repository upon publication.
\clearpage

\section*{Supplementary Information}
\setcounter{figure}{0}
\setcounter{table}{0}
\renewcommand{\figurename}{Supplementary Figure}
\renewcommand{\tablename}{Supplementary Table}

\begin{figure}[h]
  \centering
  \includegraphics[width=\textwidth]{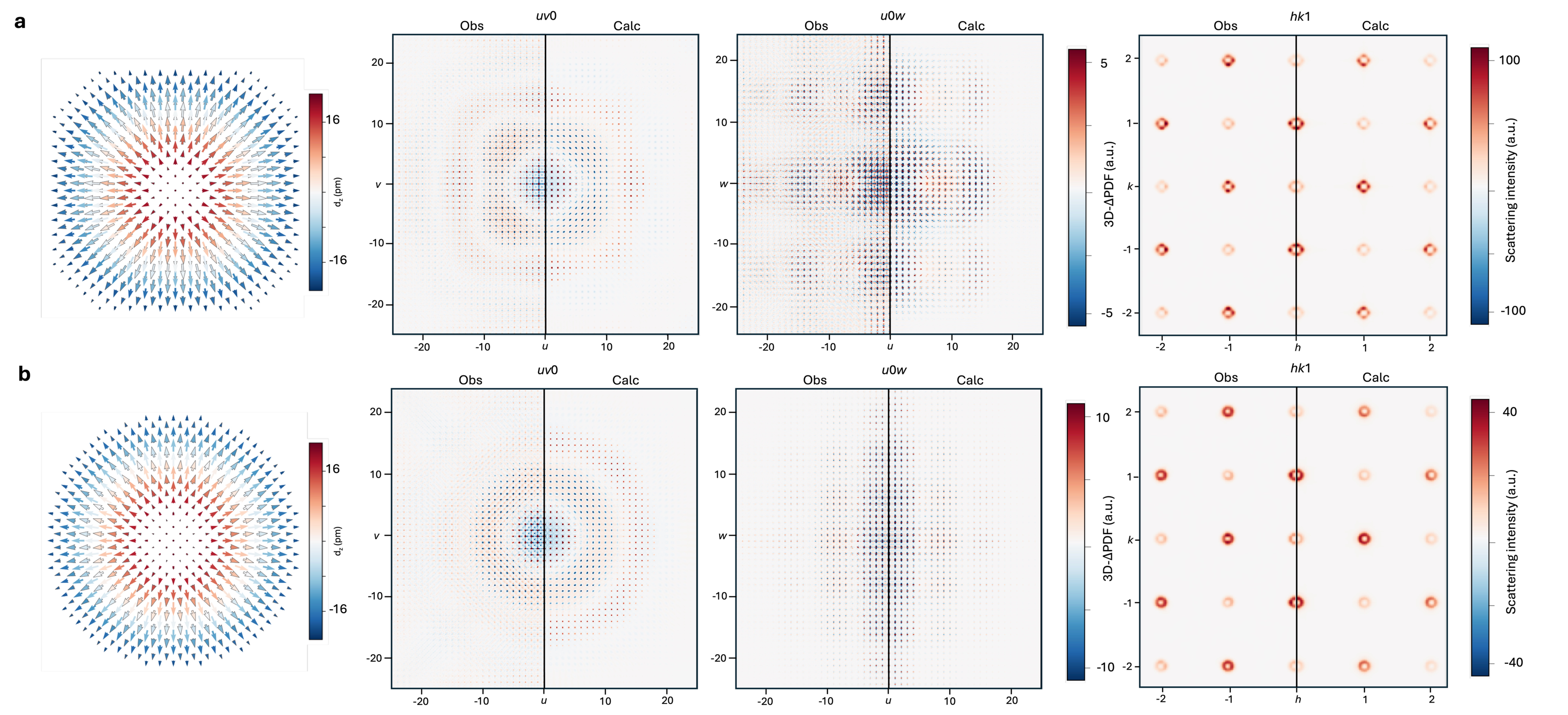}
  \caption{\textbf{Individual polar skyrmion bubble structure and their correlations in superlattices with different layer thicknesses.} \textbf{a}, $n=7$, $m=7$ superlattice: top view of a single refined bubble (left), with arrows indicating the cation displacements, coloured by their out-of-plane component $d_z$; $uv0$ and $u0w$ sections of the observed (left half) and calculated (right half) 3D-$\Delta$PDF (middle); and observed (left half) and calculated (right half) diffuse scattering in the [$hk\,l_0$] layer (right). \textbf{b}, Same for the $n=5$, $m=2$ superlattice. Calculations are for the refined lattice-gas models. Refined correlation coefficients for both superlattices are listed in Supplementary Table~\ref{tab:correlations}.}
  \label{fig:SI_samples}
\end{figure}

\begin{figure}[h]
  \centering
  \includegraphics[width=\textwidth]{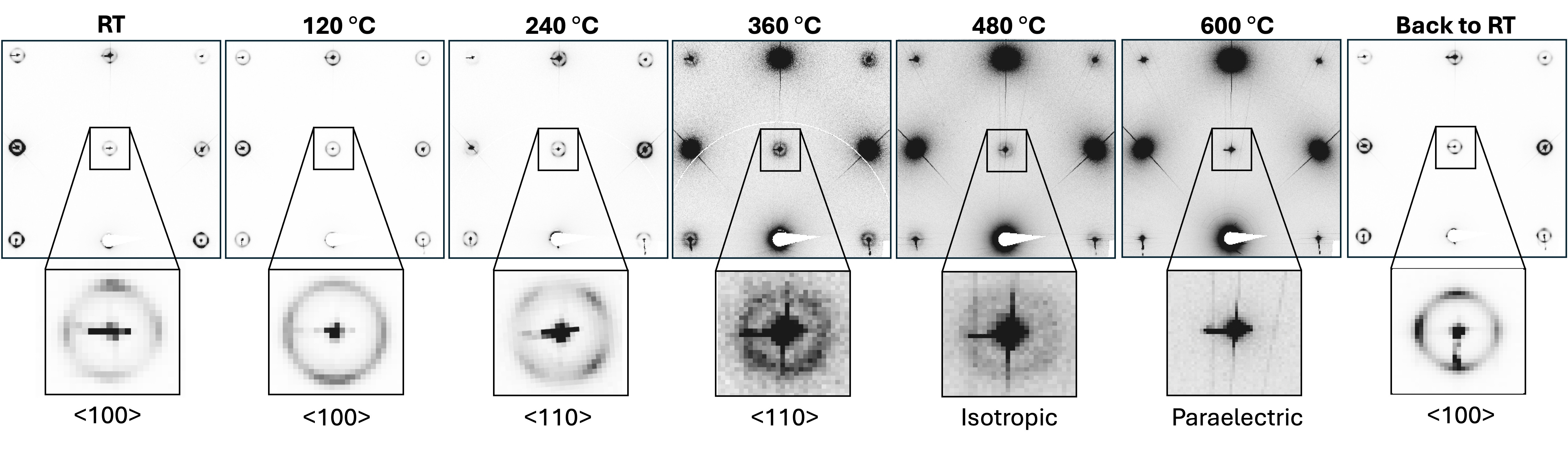}
  \caption{In-situ temperature dependence of the diffuse scattering of the $n=7$, $m=7$ superlattice. Top: $hk2$ reciprocal-space layer at room temperature, at 120, 240, 360, 480 and 600~$^\circ$C, and after cooling back to room temperature (right). Bottom: magnified views of the diffuse ring around the marked reflection. On heating, the intensity maximum around the ring moves from $\langle100\rangle$ (room temperature and 120~$^\circ$C) to $\langle110\rangle$ (240 and 360~$^\circ$C), indicating a change in the in-plane shape and arrangement of the bubbles; the ring becomes isotropic at 480~$^\circ$C and disappears at 600~$^\circ$C in the paraelectric phase. After cooling, the diffuse ring reappears, showing that the transformation is reversible. The display contrast differs between temperatures to optimize the visualization of intensity anisotropy in the diffuse ring.}
  \label{fig:SI_temperature}
\end{figure}

\begin{table}[h]
\centering
\caption{\textbf{Refined correlation coefficients and out-of-plane lattice parameters.} $\rho$ denotes the correlation coefficient between neighbouring bubbles within a layer and in the first to third neighbouring PbTiO$_3$ layers. A dash indicates a neighbour not included in the model.}
\label{tab:correlations}
\begin{tabularx}{\textwidth}{@{}l*{3}{>{\centering\arraybackslash}X}@{}}
\toprule
 & $n=3$, $m=5$ & $n=5$, $m=2$ & $n=7$, $m=7$ \\
\midrule
\multicolumn{4}{@{}l}{In plane} \\
$\rho_{100}$ & 0.23 & 0.03 & 0.20 \\
$\rho_{110}$ & 0.21 & 0 & 0.08 \\
$\rho_{210}$ & 0.08 & 0 & 0.11 \\
$\rho_{220}$ & 0.08 & 0 & \textemdash \\
\addlinespace
\multicolumn{4}{@{}l}{Out of plane} \\
$\rho_{001}$ & 0.60 & 0.62 & 0.41 \\
$\rho_{002}$ & 0.25 & 0.34 & 0.05 \\
$\rho_{003}$ & 0.06 & 0.13 & 0.01 \\
\addlinespace
\multicolumn{4}{@{}l}{Lattice parameters (\AA)} \\
$c_\mathrm{PTO}$ & 3.996 & 3.997 & 4.016 \\
$c_\mathrm{STO}$ & 3.947 & 3.948 & 3.935 \\
\bottomrule
\end{tabularx}
\end{table}

\begin{table}[h]
\centering
\caption{\textbf{Root-mean-square cation displacements.} Displacements from the centrosymmetric positions in the refined models, in pm, in plane ($x$) and out of plane ($z$).}
\label{tab:rms}
\begin{tabularx}{\textwidth}{@{}l*{3}{>{\centering\arraybackslash}X}@{}}
\toprule
 & $n=3$, $m=5$ & $n=5$, $m=2$ & $n=7$, $m=7$ \\
\midrule
\multicolumn{4}{@{}l}{PbTiO$_3$ layers} \\
Pb, $x$ & 6 & 5 & 7 \\
Pb, $z$ & 13 & 13 & 13 \\
Ti, $x$ & 5 & 4 & 6 \\
Ti, $z$ & 11 & 13 & 12 \\
\addlinespace
\multicolumn{4}{@{}l}{SrTiO$_3$ layers} \\
Sr, $x$ & 6 & 0 & 3 \\
Sr, $z$ & 7 & 4 & 2 \\
Ti, $x$ & 4 & 0 & 2 \\
Ti, $z$ & 5 & 4 & 5 \\
\bottomrule
\end{tabularx}
\end{table}

\end{document}